# Epitaxial growth and characterization of multi-layer site-controlled InGaAs quantum dots based on the buried stressor method

Imad Limame,[1] Ching-Wen Shih,[1] Alexej Koltchanov,[1] Fabian Heisinger,[1] Felix Nippert,[1] Moritz Plattner,[1] Johannes Schall,[1] Markus R. Wagner,[1,2] Sven Rodt,[1] Petr Klenovsky,[3,4,b)] and Stephan Reitzenstein[1,a)]

[1]*Institute for Solid State Physics, Technical University of Berlin, Hardenbergstraße 36, D-10623 Berlin, Germany*

[2]*Paul-Drude-Institut für Festkörperelektronik, Leibniz-Institut im Forschungsverbund Berlin e.V., 10117 Berlin, Germany*

[3]*Department of Condensed Matter Physics, Masaryk University, Kotlářská 267/2, 611 37 Brno, Czech Republic*

[4]*Czech Metrology Institute, Okružní 31, 63800 Brno, Czech Republic*

We report on the epitaxial growth, theoretical modeling, and structural as well as optical investigation of multi-layer, site-controlled quantum dots fabricated using the buried stressor method. This advanced growth technique utilizes the strain from a partially oxidized AlAs layer to induce site-selective nucleation of InGaAs quantum dots. By implementing strain-induced spectral nano-engineering, we achieve separation in emission energy by about 150 meV of positioned and non-positioned quantum dots and a local increase of the emitter density in a single layer. Furthermore, we achieve a threefold increase of the optical intensity and reduce the inhomogeneous broadening of the ensemble emission by 20% via stacking three layers of site-controlled emitters, which is particularly valuable for using the SCQDs in microlaser applications. Moreover, we obtain direct control over emission properties by adjusting the growth and fabrication parameters. Our optimization of site-controlled growth of quantum dots enables the development of photonic devices with enhanced light-matter interaction and microlasers with increased confinement factor and spontaneous emission coupling efficiency.

---

a) Author to whom correspondence should be addressed. Electronic mail: stephan.reitzenstein@physik.tu-berlin.de

b) Author to whom correspondence should be addressed. Electronic mail: klenovsky@physics.muni.cz

I. **INTRODUCTION**

The scalability of nanophotonic systems is key to unlocking their full potential and realizing their widespread applications in various fields, such as optoelectronics and emerging photonic quantum information technologies based on single quantum emitters[1]. For instance, quantum computing and communication, sensing, and quantum simulations of complex systems in chemistry, finance, and medicine development would benefit from scalable nanofabrication concepts[2–9]. Furthermore, the emission properties of optoelectronic devices such as cavity-enhanced microlasers will also be improved by a controlled integration of quasi-zero-dimensional gain centers[10]. In this regard, the deterministic fabrication of nanophotonic devices with a controlled number and position of semiconductor quantum dots (QDs) is an important step forward in achieving that goal. In the last two decades, innovative methods to simultaneously control the position of QDs have been developed, including the site-selective growth of QDs on etched nanoholes, and on pre-patterned substrates[11–13]. Another attractive growth approach for site-controlled QD is the buried stressor method. This advanced technique leads to high-quality, defect-free, and spectrally nano-engineered QDs, all in conjunction with not only precise control over the position but also the number of QDs. These properties make it a compelling technique for the deterministic fabrication of nanophotonic and quantum photonic devices[14,15].

As the name suggests, in our proposed growth method a stressor layer below the growth surface is used to control the nucleation site, and local number of the QDs[14,16]. This allows for the deterministic fabrication of single-photon sources (SPSs), which we have scaled up to an array of SPSs[15], and micropillar lasers with high confinement factor ($\Gamma$-factor), spontaneous emission coupling factor ($\beta$-factor), and low threshold power[17]. However, despite its proven potential, systematic studies on the employment of the buried-stressor method for the optimization of site-controlled QDs (SCQDs) for usage as a gain medium in microlasers are still lacking. While less than 20 SCQDs integrated in micropillar lasers have been demonstrated[18], there has been no systematic report on the optimization of the optical gain medium through QD number increase by stacking multiple layers of SCQDs, or strain engineering in combination with the growth parameter tuning so far.

Here, we report on the investigation of the buried stressor growth method to precisely engineer the QD gain medium for increasing the $\Gamma$- and $\beta$-factors, while at the same time reducing the absorption losses due to spatially and spectrally non-positioned QDs. This paves the way to the energy-efficient operation of those devices, which is of great interest, particularly for applications requiring a large number of devices, such as photonic reservoir computing, where for usage as a nanophotonics platform hundreds of microlasers need to be pumped above threshold and optically coupled[19,20]. Furthermore, we show that by using strain engineering not only for site control but also for spectral tuning, the technique could serve as an alternative approach for growing high-quality QDs emitting prospectively in the telecom O-band, to complement existing non-positioning methods such as strain-reducing layers (SRLs) and metamorphic buffer layers (MB)[21,22].

The buried stressor technique takes advantage of the strain induced by a partially oxidized AlAs layer, stemming from a lattice constant difference between AlAs (5.660 Å) and $Al_2O_3$ (4.785 Å). The resulting tensile strain at the center of the etched mesa protrudes to the surface, where it leads to the accumulation of indium atoms at the most tensile strained positions (largest lattice constant) above the aperture, which results in the site-controlled nucleation of QDs[14]. In addition, control over the emission wavelength is achieved by optimizing the growth interruption time. In fact, red-shifted SCQDs with respect to the emission of QDs in sample surface areas not affected by the buried stressor-induced tensile strain, are achieved for long growth interruption times in the range of 40 to 60 seconds. The spectral separation of the two types of QDs allows us to grow higher QD densities without increasing the optical absorption by spatially and spectrally detuned non-positioned QDs. To further increase the optical gain for laser applications, we stack multiple layers of SCQD (ML-SCQDs). Compared to conventional SK QDs, the SCQDs stacking leads to a higher optical gain, and a better spectral homogeneity, as well as position control of QD formation [23].

Our samples were grown using metal-organic chemical vapor deposition (MOCVD). Atomic force microscopy (AFM) was employed to investigate the strain profile at the QD growth surface and to measure the QD density. The microscopic analysis is complemented by Raman spectroscopy above the aperture, which is used to monitor the strain profile by measuring the strain-induced spectral shift of the longitudinal optical phonon (LO) line of GaAs. Moreover, we applied the finite-difference method to simulate the strength and shape of induced stress and its effect on the SCQDs using continuum elasticity theory[24]. During the lateral oxidization, the aperture layer from the sample edges changes the material composition from AlAs, which has a zinc blende crystal structure, to $Al_2O_3$ which has a triclinic structure. That complicates the continuum elasticity computation, which would either need to be slightly redefined or atomistic methods like molecular dynamics should be used[25,26]. Nevertheless, since it is questionable whether the oxidized AlAs material in the aperture layer has the same crystal structure and all properties as $Al_2O_3$ and for the sake of simplicity, we decided to deliberately consider instead an artificial material with parameters the same as AlAs, but with lattice parameter of 4.785 Å, corresponding to a-axis of triclinic crystal (the c-axis in $Al_2O_3$ has lattice parameter of 12.99 Å). Finally, high-resolution cathodoluminescence (CL) was utilized to characterize the emission properties of the emitters. Our approach allows us to shift the energy of the site-controlled emitters compared to the non-positioned QDs, enabling, thus, the growth of SCQD with high density without compromising the site-controlled nature of the buried stressor technique. Moreover, compared to non-positioned QDs we enhance the optical gain threefold while reducing the inhomogeneous broadening of the ensemble by stacking three layers of SCQDs. Finally, we present a new approach to obtain direct control over emission properties by engineering the fabrication parameters.

## II. RESULTS AND DISCUSSION

This section is structured as follows: in section A. we discuss the growth and emission properties of a single layer of SCQDs, in section B. we focus on the emission optical intensity of a stack of SCQDs, and in section C. we study the methods of strain engineering of QD emission wavelength.

### A. Optical and strain analysis of SCQDs

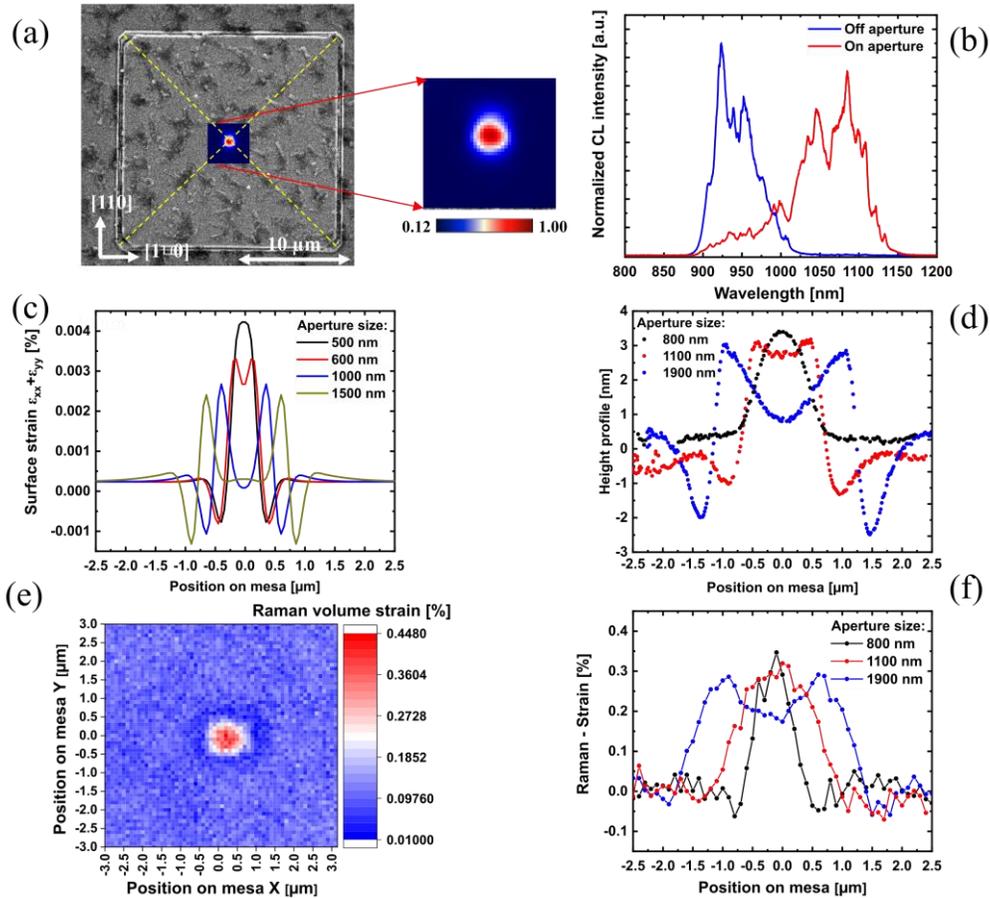

FIG. 1. Optical and strain analysis of SCQDs. (a) Low temperature (20 K) CL intensity map of SCQD emission ranging from 1020 to 1120 nm is overlaid with an SEM image of the quadratic mesa, including the crystallographic orientation of the mesa edges. The high position accuracy of the growth technique is visually demonstrated by the presence of two yellow dotted lines. In the right

inset, we show a zoom-in image taken at the center of the mesa (b) The CL spectrum is shown in red and blue for the center and off-center regions of the mesa, respectively. (c) A continuum elasticity theory simulation depicting the surface strain profile $\varepsilon_{xx} + \varepsilon_{yy}$ along the cross-section of the mesa, as a function of the aperture size. $\varepsilon_{xx} + \varepsilon_{yy} = 0$ corresponds to unstrained GaAs. (d) AFM height profile of the aperture for various aperture sizes. (e) Strain taken at the center of the mesa, obtained by frequency conversion of the Raman scan of the LO-phonon band of GaAs. (f) A line cut to the center of the mesa of the Raman map in (e) for three different aperture sizes (800, 1100, and 1900 nm).

Figure 1(a) shows a CL intensity color map of a single layer of SCQDs with a growth interruption time of 40 s in the wavelength range between 1000 and 1150 nm overlapped with an SEM image of the mesa, demonstrating high position accuracy (below 100 nm) of the buried stressor method. Note, that sub-µm alignment accuracy is of great importance for applications that require the deterministic integration of single SCQDs or small ensembles of those emitters at the center of a nanophotonics structure[18,27]. While SCQDs based on nanohole arrays and inverted pyramids feature an alignment accuracy in the tens of nm and are suitable for integration into photonic crystal cavities, buried stressor SCQDs, with usually better optical properties, are more suitable for integration in high Q-factor cavities with relaxed mode-matching requirements on the order of a few 100 nm[28,29]. CL emission of on- and off-center from the mesa shown in Fig. 1(a) is displayed in Fig. 1(b). The SCQDs have a maximum emission wavelength of 1070 nm (red curve in Fig. 1(b)) and exhibit a strongly redshifted emission in comparison with the non-positioned QDs of 950 nm (blue curve in Fig. 1(b)). For comparison, the wetting layer emission occurs at 920 nm (see Supplementary Information (SI), Fig. S2). The shift of the emission wavelength is achieved by higher localized tensile surface strain above the aperture combined with a longer growth interruption (of 40 s) time (see section C for details). The observed magnitude of the redshift is compatible with an increased indium content in the SCQDs as well as with larger QDs. We interpret the observed behavior by buried stressor-induced migration of the indium atoms to the most tensile strained position directly above the aperture, resulting in a higher indium concentration in the center of the mesa. This together with a higher InGaAs growth rate, hence, leads to an increase in density, size, and indium content of SCQDs compared to non-positioned QDs, thus, contributing to the observed redshift of SCQD emission. In this process, the growth interruption time enables us to achieve an equilibrium that favors more red-shifted emitters in contrast to non-positioned QDs grown on compressively strained GaAs surface, which will be discussed in the following section. Intriguing, apart from the achieved spectral redshift, the peak intensity of the SCQDs is maintained compared to the non-positioned QDs. We attribute that to the high quality of the growth surface and low defect density compared to other methods for redshifting QD emission using a SRL or a MB.

To obtain better insight into the buried stressor growth of SCQDs, the surface strain distribution at the growth surface was modeled using continuum elasticity theory. Simulations of the strain profile at the GaAs surface versus the position on the mesa for different aperture sizes are depicted in Fig. 1(c). The difference in lattice constants between AlAs and $Al_2O_3$ results in lateral strain, which propagates in growth direction to the surface (see the SI, Fig. S3 (a) and (b)), where the InGaAs wetting layer is deposited. The buried stressor creates tensile and compressive stress with surface strain extrema of 0.4% and -0.12%, respectively. The strain profile depends in this configuration solely on the size of the aperture as seen in Fig. 1(c), whereas the strain magnitude depends on the size of the aperture as can be observed in Fig. 1(c), the stressor layer thickness, and the aperture to surface distance (see SI, Fig. S5 (a)). Corresponding AFM measurements in Fig. 1(d) show a smooth height profile after the growth of a single uncapped QD layer for different aperture sizes. Here, the migration of indium atoms from the compressive to the tensile position leads to a change in growth rate directly above and surrounding the aperture. As shown in Fig. 1(d) this results in a thickness difference in growth direction, whereas the unstrained GaAs on the mesa remains unchanged. We denote the height difference between most tensile positions located above the aperture and the GaAs surface as $\Delta_t$, a parameter that characterizes the magnitude of the strain-induced growth rate change above the stressor. As predicted by simulations, the smallest aperture exhibits the highest $\Delta_t$ due to an increase in surface strain with a smaller aperture size. As the compressively strained position is depleted of indium, no QD growth is observed in related areas in the AFM images (see the SI, Fig. S. 4). The indium atoms that migrate to the nearby tensile strained positions above the aperture, resulting in a higher SCQDs density, as seen in AFM images (see the SI, Fig. S. 4) with higher indium content, explaining the redshift of the SCQD emission in Fig. 1(b). The lack of QD growth around the aperture can also be observed in the CL maps (see white ring surrounding the aperture in Fig. 1(a) right), as the luminescence from the surrounding area of the aperture is weak to non-existent.

The strain distribution is monitored by measuring the Raman emission in the surrounding GaAs above the unoxidized AlAs layer. For this purpose, we performed systematic Raman spectroscopy measurements. A corresponding color map of the GaAs LO-phonon Raman band recomputed to strain for a mesa with an aperture size of approximately 1000 nm is shown in Fig. 1(e). The stressor-induced surface strain change is associated with a change in the bond lengths, resulting in a shift of the GaAs LO-phonon band, in this case resulting in a Raman shift of approximately 1.7 cm$^{-1}$, corresponding to a strain magnitude of $(0.35 \pm 0.001)$%. The induced strain is evident in the line scans displayed in Fig. 1(f) for three different-sized apertures. The noteworthy aspect here is the very good quantitative agreement between the computed strain (approximately 0.3%) and the measured strain ((0.35 ± 0.001)%), as evident in Fig. 1(c) and (f). This substantiates the predictive nature of the simulation.

## B. Optical properties of multi-layer SCQDs

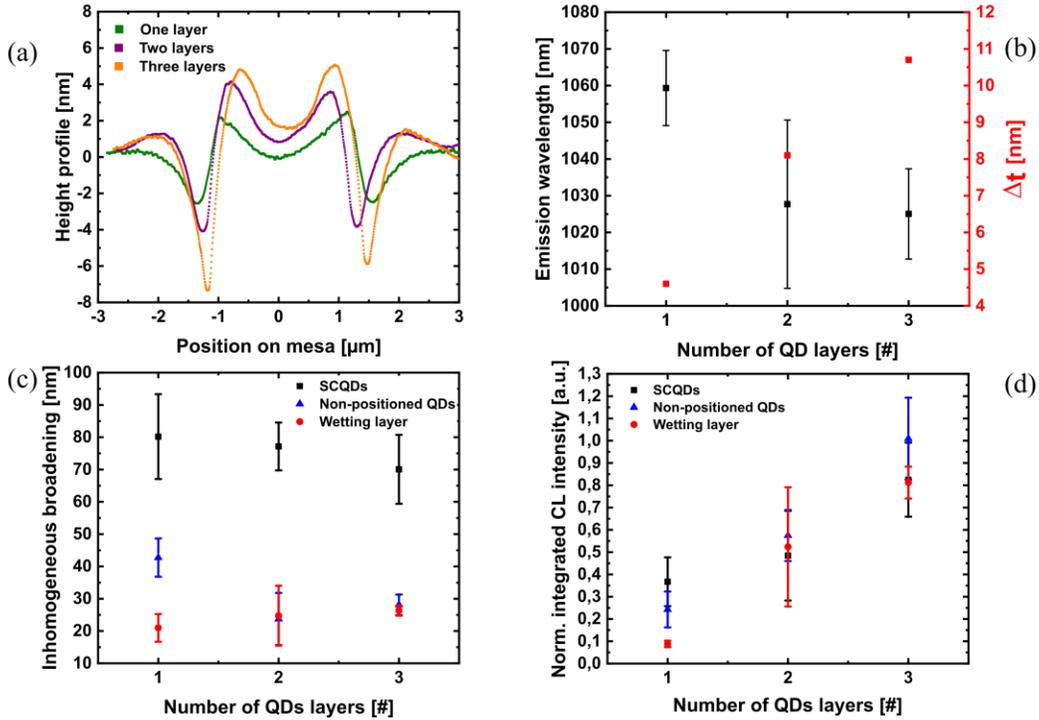

FIG. 2. Properties of multi-layer SCQDs. (a) Line scan of AFM measurements, presenting the height profile for a varying number of SCQD layers at the center of the mesa with similar aperture sizes (approx. 2 µm), relative to the GaAs surface. (b) The black data points represent the emission wavelength of the SCQDs relative to the number of deposited emitter layers. The red data points indicate the height difference between the most tensile and compressive positions in the center of the mesa. (c) The CL emission spectra exhibit inhomogeneous broadening for different numbers of stacked layers, including the wetting layer (red), non-positioned QDs (blue), and SCQDs (black). (d) The integrated CL intensity is plotted against the number of grown layers, revealing enhanced intensity with increased layer stacking.

In this section, we explore the impact of stacking multiple QD layers on both the surface morphology and optical properties, particularly the inhomogeneous broadening, optical intensity enhancement, and shift in emission wavelength. In Fig. 2(a), we present the AFM height profiles at the center of the mesa for one, two, and three layers of SCQD structures with similar aperture sizes (approx. 2 µm). We observe that stacking of multiple QD layers does not alter the overall shape of the surface strain profile. However, the height difference $\Delta_t$ between the most tensile and compressive positions increases systematically with the number of deposited QD layers, as indicated by the red data points in Fig. 2(b). As discussed earlier, the shape of the strain profile is solely influenced by the size of the oxide aperture. The gradual increase of $\Delta_t$ with the number of QD layers is attributed to the larger tensile strain due to the increased distance between the aperture and SCQD layer in addition to the accumulation of the indium, which persists across the multiple layers. Therefore, the growth of multiple layers and the stress induced by the QDs do not seem to modify the surface strain profile significantly. Since the QD layers are separated by a 20 nm thick GaAs spacer layer, each subsequent layer experiences a larger distance from the aperture. Consequently, there is an approximately 0.7% change in the absolute surface strain between the first and the third layer. However, we do not expect any significant effects on the formation of SCQDs, which is confirmed by the constant QD density distribution observed across the layers (see the SI, Fig. S5).

To evaluate the optical properties of the QDs we fit the spectra obtained from CL measurements, as shown in Fig. 1(b), by a Gaussian function, across multiple mesas and variations in the number of layers. From that analysis, we obtain the emission wavelength, integrated intensity, and inhomogeneous broadening, all correlated with the number of deposited QD layers. In Fig. 2(b), we present the emission wavelength of the SCQDs. Notably, we observed a distinct blue shift in the emission when two layers were stacked, reducing the wavelength from (1059 ± 10) nm for a single layer to (1027 ± 22) nm for the two-layer structure. This shift can be attributed to a decrease in the average height of the QDs in the second layer, resulting in enhanced quantum confinement, analogous to standard SK QDs[23]. Furthermore, Fig. 2(c) demonstrates a similar trend regarding the inhomogeneous broadening. While the broadening of the wetting emission increases slightly with the number of layers, the linewidth of the non-positioned and SCQDs decreases. Initially, the broadening of the emission for the first QD layer was found to be (80 ± 13) nm, because of the wide range of size distribution among QDs. However, as the number of layers increases, the full width at half maximum (FWHM) decreases for both non-positioned and positioned QDs to (67 ± 9) nm and (28 ± 3) nm, respectively, indicating improved uniformity of QD sizes. The dominant contributing factor to this effect is attributed to the growth of subsequent QDs above the prior QD layer induced by strain [23]. Notably, a two-fold larger FWHM of SCQDs when compared to the non-positioned emitters is observed. This can be explained by emission from unstrained QDs within the aperture due to the shape of the strain distribution, deduced from the height in Fig. 1(d). Therefore, as depicted in Fig. 1(b), we observe that the emission of the SCQDs ranges from the wetting layer at 920 nm up to 1150 nm.

In addition to the improved inhomogeneous broadening of QD ensemble emission and the blue shift in the emission energy, the stacking of SCQDs results in a nearly linear increase in integrated CL intensity, as shown in Fig.

2(d). That follows from the increase of QDs density due to stacking. The optical intensity increase observed in Fig. 2(d) exhibits a lack of saturation, which is expected due to defect formation and deteriorating quality for a high number of layers, allowing for further SCQD stacking beyond three layers shown here. Overall, our results show that the combination of emission shift and SCQD stacking, paves the way to engineering high optical gain micropillar lasers with a strong overlap between the gain medium and the optical mode.

### C. Strain-induced redshift of SCQDs emission

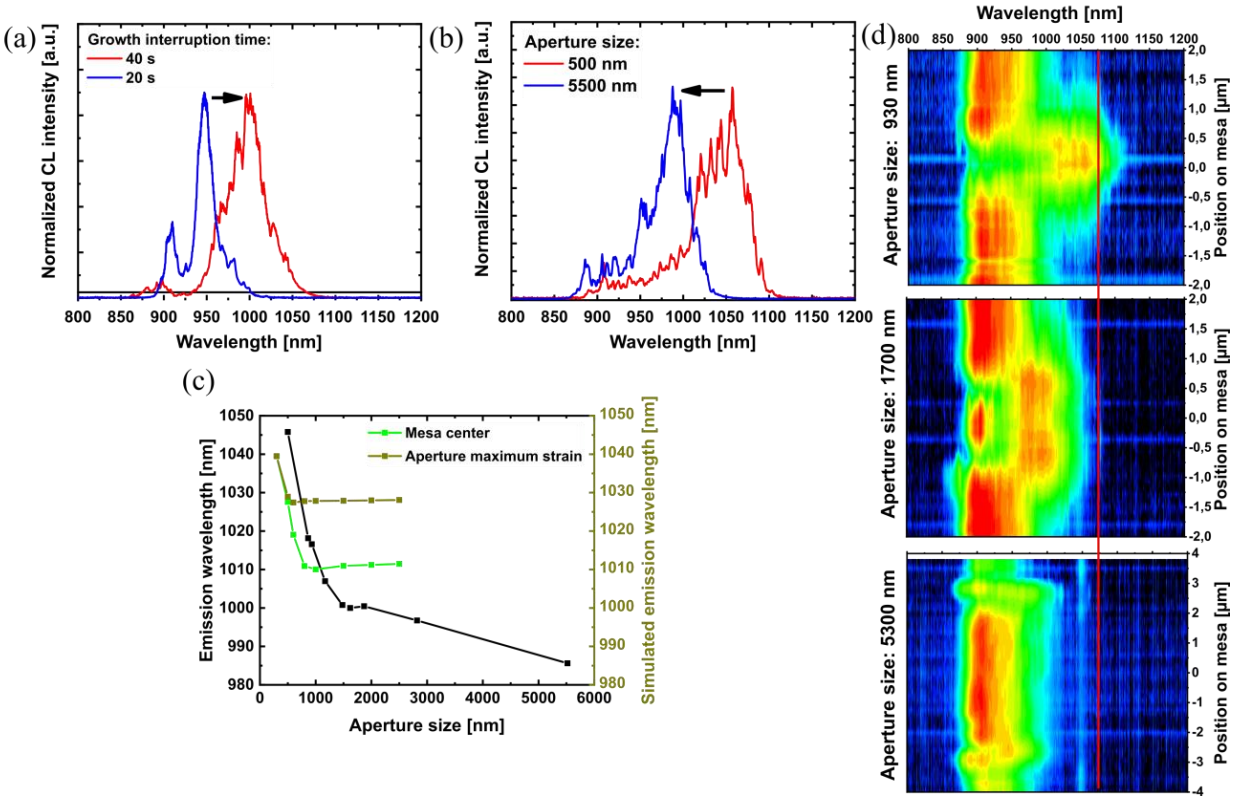

FIG. 3. Strain-engineering the emission wavelength of SCQDs. (a) CL spectra of two similar structures containing SCQDs are shown. The blue spectrum corresponds to QDs obtained with a growth interruption time of 20 seconds, while the red spectrum corresponds to a 40-second interruption. (b) The CL emission from a structure with 40 seconds of growth interruption time for two aperture sizes. Increasing the size of the aperture reduces the tensile surface strain, leading to a blue shift in the emission. (c) The experimental peak position of the emission for the SCQDs (black data points), and the results of the continuum elasticity theory in conjunction with k.p and CI simulations performed to determine the relationship between the exciton emission wavelength against the aperture size. The theory results are provided for the surface strain at the center of mesa (green squares) and for maximum surface strain on the aperture (khaki squares). The experimental values and the theoretical prediction are in good quantitative agreement. Deviations are explained by simulated QD size and structural properties not exactly matching that in the experiment. (d) Examples of CL line scan displaying the emission properties in comparison to the strain profile above the aperture for different sizes is shown.

The buried stressor growth approach takes advantage of the strain engineering of the SCQDs' optical properties compared to those of non-positioned QDs to achieve higher local emitter density and higher optical gain without compromising the site control of the buried stressor method. Interestingly, it also provides an attractive opportunity to redshift the emission of SCQD, which eventually might reach the telecom O-band. For example, consider the strain magnitude or, more precisely, the shift in the GaAs LO band associated with an aperture size of 1000 nm. The corresponding Raman frequency shift is approximately 1.65 cm$^{-1}$, a value comparable to that of an In$_{0.15}$Ga$_{0.85}$As MB with a thickness ranging from approximately 40 nm to 50 nm[30]. To demonstrate the strain-engineered redshifting of SCQD emission, we utilize the buried stressor-induced strain profile above the aperture. Here, the additional tensile strain in the center of the mesa leads to energetically lower QD equilibrium compared to the unstrained surface. Secondly, by increasing the growth interruption time, we allow for the SCQDs to accumulate more Indium atoms, resulting in bigger dots with increased Indium concentration. That can be observed from CL as a strong red shift of the emission as seen in Fig. 3(a). The blue spectrum represents a sample prepared with a growth interruption time after the deposition of InGaAs of 20 seconds, while the red curve is associated with a growth interruption of 40 seconds. We can see a clear red shift of the emission wavelength of about 70 nm. As described above, this is attributed to the higher tensile strain above the aperture leading to larger, more indium-containing SCQDs. Additionally, we expect the tensile strain induced by the aperture to also further red shift the emission. This is also confirmed by our exciton calculations performed using a combination of the eight-band k.p method with the configuration interaction correction (CI). Note that the surface strain obtained from continuum elasticity theory is added to the as-grown QD strain during the k.p computation[31]. See the SI, Fig. S6 (a), where the computed applied tensile strain leads to a red shift of the QD emission. The increase of the aperture size results in a weaker lateral strain between AlAs and Al$_2$O$_3$, which in turn decreases the stress propagation to the surface. Moreover, altering the aperture size results in a change in the strain profile's general shape. It shows a transition from a single peak of tensile strain at the center of the mesa to two peaks in the cross-section view, resulting in a circular or square aperture shape, as observed in the AFM measurements. The increased magnitude of the tensile strain at the center of the mesa due to the reduction of the aperture size results in larger QDs, with increased indium content, leading to a red shift in the emission of the SCQDs, as depicted in Fig. 3(b), where the emission wavelength shifts from 1050 nm for an aperture size of about 500 nm to approximately 980 nm for a size of 5500 nm. In Fig. 3(c) the emission wavelength is plotted against the aperture size measured by CL spectroscopy and simulated using continuum elasticity theory combined with eight-band k.p and CI, respectively. In both graphs, the emission wavelength is highest for the smallest aperture size, then drops rapidly with increasing aperture size. That is due to the reduction in surface strain with a bigger unoxidized AlAs region. Consequently, the migration of indium atoms and an increase in growth rate above the aperture occurs. The aforementioned dependence of strain magnitude, its distribution, and its resulting impact on the emission properties of the SCQDs can be vividly seen in the CL line scans of apertures with varying sizes, as presented in Fig. 3(d). Apertures of 530 nm, 1700 nm, and 5300 nm in width result in peak emission wavelengths of 1070 nm, 1010 nm, and 985 nm, respectively. This precise control of emitter distribution by adjusting the mesa size sets the stage for integrated single photon sources (SPS) with small and localized strain (apertures below 1 µm), microlasers featuring larger apertures (up to 1.5 µm), as well as whispering gallery microcavity lasers (above 2 µm apertures) utilizing the circular distribution of the gain medium. In order to shift the emission towards the telecom O-band, we found that a strategy based on combination of thicker AlAs stressor layer and top GaAs spacer, lead to an increased tensile surface strain, and associated increase of Indium content in QDs should be adopted.

### III. CONCLUSION

In summary, we introduced a novel approach based on the buried stressor technique, to achieve a redshift of SCQDs relative to non-positioned ones. This approach enables the deposition of high-density SCQDs within a singular layer, preserving positional control and mitigating increased absorption losses associated with higher non-positioned QD densities. This is achieved while concurrently monitoring and simulating strain distribution to gain deeper insights into the underlying mechanisms. Furthermore, we demonstrate the successful stacking of three SCQD layers, resulting in a threefold increase of optical intensity and a reduction in ensemble inhomogeneous broadening. Subsequently, we investigate the strain-induced redshift of the SCQD ensemble, granting us direct control over emission properties. Combining the buried stressor technique with strain nano-engineering of emission provides control over the site, number/density, and optical properties of both ensembles and single QDs. This achievement can potentially pave the way not only for high-$\beta$ factor and low-threshold microlasers, but also for single-photon sources operating in the O-band without the need of a strain relaxing layer.

**ACKNOWLEDGMENTS**

The authors gratefully acknowledge the financial support from the Volkswagen Foundation via the project NeuroQNet2, the German Research Foundation via INST 131/795-1 320 FUGG, and the SEQUME (20FUN05) and QADeT (20IND05) from the EMPIR program cofinanced by the Participating States and from the European Union's Horizon 2020 research and innovation program. P.K was partly funded by the Institutional Subsidy for the Long-Term Conceptual Development of a Research Organization granted to the Czech Metrology Institute by the Ministry of Industry and Trade of the Czech Republic and by the project Quantum Materials for applications in sustainable technologies, CZ.02.01.01/00/22_008/0004572. The authors further acknowledge Kathrin Schatke, Prapaht Sonka, Lucas Rickert, Aris Koulas-Simos, and Maximilian Ries for their invaluable technical support and engaging scientific discussions, which greatly contributed to the development and success of this research.

## AUTHOR DECLARACTIONS

### CONFLIFT OF INTEREST

The authors have no conflicts to disclose.

### AUTHOR CONTRIBUTIONS

**Imad Limame:** conceptualization (equal), data curation (lead), formal analysis (lead), investigation (lead), methodology (lead), resources (lead), visualization (lead), writing/original draft preparation (equal), writing/review & editing (equal). **Ching-Wen Shih:** investigation (equal), methodology (equal), resources (supporting). **Alexej Koltchanov:** data curation (equal), investigation (supporting). **Fabian Heisinger:** data curation (supporting), formal analysis (equal), methodology (equal), software (equal). **Felix Nippert:** formal analysis (supporting), methodology (supporting). **Moritz Plattner:** data curation (supporting), investigation (supporting). **Johannes Schall:** methodology (supporting), software (supporting). **Markus Wagner:** supervision (supporting), validation (equal). **Sven Rodt:** methodology (supporting), supervision (supporting). **Petr Klenovsky:** funding acquisition (equal), methodology (supporting), resources (supporting), software (equal), validation (equal), visualization (equal), writing/review & editing (supporting). **Stephan Reitzenstein:** conceptualization (equal), funding acquisition (lead), project administration (lead), supervision (lead), validation (lead), writing/original draft preparation (equal), writing/review & editing (equal).

### DATA AVAILABILITY

The data that support the findings of this study are available from the corresponding author upon reasonable request.

# *Supplementary information:* Epitaxial growth and characterization of multi-layer site-controlled InGaAs quantum dots based on the buried stressor method

Imad Limame,[1] Ching-Wen Shih,[1] Alexej Koltchanov,[1] Fabian Heisinger,[1] Felix Nippert,[1] Moritz Plattner,[1] Johannes Schall,[1] Markus R. Wagner,[1,2] Sven Rodt,[1] Petr Klenovský,[3,4,b)] and Stephan Reitzenstein[1,a)]

[1]*Institute for Solid State Physics, Technical University of Berlin, Hardenbergstraße 36, D-10623 Berlin, Germany*

[2]*Paul-Drude-Institut für Festkörperelektronik, Leibniz-Institut im Forschungsverbund Berlin e.V., 10117 Berlin, Germany*

[3]*Department of Condensed Matter Physics, Masaryk University, Kotlářská 267/2, 611 37 Brno, Czech Republic*

[4]*Czech Metrology Institute, Okružní 31, 63800 Brno, Czech Republic*

a) Author to whom correspondence should be addressed. Electronic mail: stephan.reitzenstein@physik.tu-berlin.de

b) Author to whom correspondence should be addressed. Electronic mail: klenovsky@physics.muni.cz

In this supplemental information, we provide details on the fabrication, characterization, and simulation methods in section A of the main text, followed by additional optical, structural, and simulation results.

## 1. Fabrication, characterization, and simulation methods

The fabrication of site-controlled quantum dots (SCQDs) based on the buried stressor approach involves multiple steps of growth and processing[1]. As depicted in Fig. S1(a), the first step is the metal-organic chemical vapor deposition (MOCVD) growth of a template structure on a 400 µm thick n-doped, GaAs:Si (001) 2" wafer. A 300 nm GaAs buffer layer is deposited at a temperature of 700 °C and a reactor pressure of 100 mbar to obtain a high-quality epitaxial surface. Subsequently, 15 pairs of λ/4 $Al_{0.9}Ga_{0.1}As$/GaAs with nominal thicknesses of 77.8 and 66.5 nm are grown, forming a back side distributed Bragg reflector (DBR) to enhance the photon extraction efficiency. This mirror is followed by a 50 nm grading layer that serves as a transition from GaAs to 30 nm of binary AlAs, which later functions as the buried stressor. Finally, an inverted grading layer is grown, capped with an 80 nm GaAs layer to prevent any undesirable oxidation of the surface.

The sample is subsequently cleaved into four quarter wafer pieces, and each is processed separately. The structure is spin-coated with AZ 701MIR photoresist at 3000 rpm and baked for one minute at 100 °C under nitrogen flow. The ultraviolet (UV) lithography is then performed to define arrays of square mesas with side lengths ranging between 20 and 21 µm with 63 nm step size into the photoresist, with the mesa edge aligned to the (110) crystal plane. The alignment choice depends on the desired shape of the final aperture, as the oxidation speed depends on the lattice orientation[2]. In our case, this results in a diamond to circular-shaped mesa depending on the size of the unoxidized AlAs layer, i.e. the aperture. Afterward, a one-minute post-exposure baking at 110 °C is necessary to harden the photoresist. The structures are then developed using AZ 726 MIF. Reactive ion etching, utilizing an inductively coupled plasma (ICP-RIE), is employed to transfer the pattern into the III/V semiconductor. The sample is dry etched down to the tenth mirror pair of the DBR to form the intended square mesa structures. A secondary electron microscope (SEM) image of a processed mesa can be seen in Fig. S1 (d). After the etching step, the AlAs layer is exposed. The selective wet oxidation of the pure AlAs layer is performed in a vacuum furnace, using a mixture of nitrogen (0.8 liter/minute) and water vapor (0,8 liter/minute) as the process atmosphere at a pressure of 50 mbar and a temperature of 420 °C. This process is depicted in Fig. S1 (b). The total oxidation of a 20 µm mesa takes approximately 13 minutes. The size of the aperture is in-situ monitored with an optical microscope. The process is halted when the desired aperture size is reached by flooding the chamber with nitrogen gas (3 liters/minute) and turning off the heaters to initiate the cooling process. The final structure consists of a small, and size-controlled unoxidized AlAs aperture in the center of the mesa laterally surrounded by $Al_2O_3$, as seen in Fig. S1 (e). This process is highly selective and the DBR layers are not affected as the concentration of Al is lower than 95% [2]. The process and final mesa structure are illustrated in Fig. S1 (b).

The final step of the SCQDs fabrication starts with the meticulous cleaning of the patterned sample and removal of surface oxides by immersing the sample for 30 seconds in a 75% sulfuric acid solution. Afterward, the second epitaxial step commences with a 50 nm thick GaAs layer as depicted in Fig. S1 (c), deposited at a temperature of 700 °C. Due to the high temperature and unetched surface, high-quality QD growth is ensured. The total distance between the stressor and QDs layer is 130 nm. A 1.8 monolayer thick $In_{0.63}Ga_{0.37}As$ wetting layer is then grown at a temperature of approximately 500 °C. Following a growth interruption between 20 and 40 seconds, a 1.2 nm thick layer of GaAs is deposited before increasing the temperature to 615 °C to desorb the remaining, uncapped indium on the surface and grow an 18 nm GaAs barrier to prevent tunnel coupling between the SCQD layers. This process is repeated for the growth of additional QD layers. For atomic force microscopy (AFM) studies, the QDs are capped with 1.2 nm GaAs at 500 °C, in contrast to 50 nm for samples intended for optical studies using cathodoluminescence (CL) spectroscopy.

All growth rates and concentrations were calibrated using high-resolution X-ray diffraction (HR-XRD). Following a growth interruption between 20 and 40 seconds, a 1.2 nm GaAs is deposited, before increasing the temperature to 615 °C to desorb the remaining, uncapped indium on the surface and grow an 18 nm thick GaAs barrier to prevent tunnel coupling between the SCQD layers. This growth process is repeated for the growth of additional QD layers.

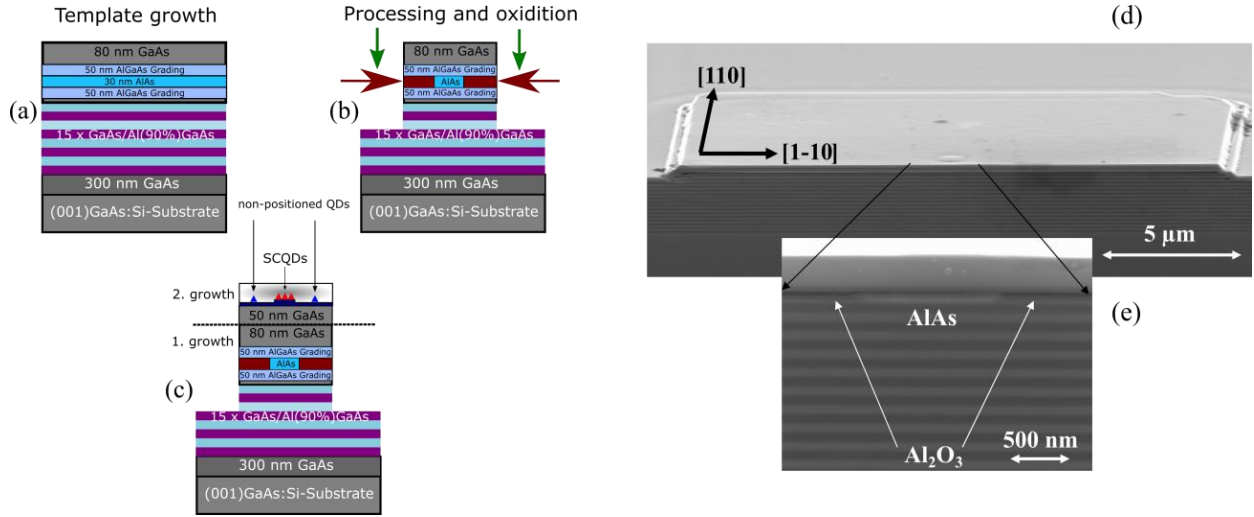

FIG. S1. (a) The initial step in epitaxy involves growing a template structure, which includes a 300 nm buffer layer, 15 pairs of GaAs/Al$_{0.9}$GaAs DBR, a 30 nm AlAs stressor layer, and an 80 nm GaAs cap. (b) Following the UV lithography step, selective wet oxidation at 420°C, results in a precisely controlled unoxidized AlAs aperture in the center of the mesa. (c) The second epitaxy step commences with a high-quality 50 nm thick GaAs layer. The strain emanating from the aperture induces the accumulation of indium atoms and enables site-controlled nucleation of QDs above the aperture. (d) A cross-section SEM image of a cleaved mesa reveals a visible height profile in the center of the square mesa. (e) An enlarged image provides a closer look at both the oxidized and unoxidized regions of the AlAs layer.

In Fig. S1 (d) a cross-sectional secondary electron microscope (SEM) image of a cleaved mesa can be seen. In the center of the structure, the indium migration induced by the strain of the aperture results in higher and lower thicknesses at the tensile and compressive strained position, respectively. The zoom-in image in Fig. S1 (e) shows the contrast between the Al$_2$O$_3$ and AlAs layers.

The optical studies were conducted in a helium-cooled cathodoluminescence (CL) setup at a temperature of 20 K. The electron excitation beam with a 2 nm focus size and an acceleration voltage of 5 kV with an aperture of 30 µm was used, resulting in a current of 50 pA. A high NA parabolic mirror is used to collect the emitted luminescence. The emission is then directed on a monochromator with a slit size of 100 µm and a grating with 300 lines/mm. First, an overview 20 µm x 20 µm CL map is recorded with 500 nm pixels to evaluate the precision of the site-control accuracy.

Then, a high-resolution map with 100 nm resolution to study the shape of the aperture and optical properties are measured. Both measurements are recorded at 200 ms integration time per pixel.

Raman spectroscopy was employed to monitor the distribution of strain using a Thermo Scientific DXR3xi imaging microscope. To achieve that, a 532 nm green diode-pumped solid-state laser with an excitation power of 1.7 mW was focused onto the sample through a 100x objective with an NA of 0.9. A scanning area of 6 µm x 6 µm was mapped, centered around the aperture, with a step size of 100 nm. To enhance spatial resolution, a 25 µm pinhole was introduced. A grating with a resolution of 2 cm$^{-1}$ was used, and a CCD was utilized for spectrum recording. In each measurement, a Lorentzian function was employed to fit the GaAs LO line (292 cm$^{-1}$), and the resulting Raman frequency was extracted. The strain values were subsequently calculated based on this preview works[3–5].

We theoretically calculated the surface biaxial strain induced by the aperture using the continuum elasticity theory, which relies on minimizing the total strain energy in the whole simulated structure. During the oxidization, the aperture layer from the sample edges changes the material composition from AlAs, which has a zinc blende crystal structure, to $Al_2O_3$ which has a triclinic structure. That complicates the continuum elasticity computation, which would either need to be redefined or atomistic methods like molecular dynamics should be used[6]. Nevertheless, since it is questionable whether the oxidized AlAs material in the aperture layer has the same crystal structure and all properties as $Al_2O_3$ and for the sake of simplicity we deliberately considered in the simulations instead an artificial material with the parameters the same as for AlAs, but with lattice parameter of 4.785 Å, corresponding to a-axis of triclinic crystal (the c-axis in $Al_2O_3$ has lattice parameter of 12.99 Å).

For the computation of the excitonic energy, we used a combination of eight-band k.p envelope function approximation with corrections due to the Coulomb interaction between electrons and holes by configuration interaction method (CI) with a single-particle basis of two-electron and two-hole single-particle ground states[7]. We further note, that the effect of aperture-induced biaxial strain was added to the QD strain, and the single-particle eigenenergies and eigenfunctions were computed thereafter by eight-band k.p and CI[8].

## 2. Experimental and simulation results

Fig. S2 (a) and (b) depict an SEM image overlapped with a CL intensity map at the center of the mesa for the wetting layer emission range (885-915 nm) and non-positioned QDs (920-980 nm), respectively. In both instances, we observe a reduced intensity within the aforementioned wavelength ranges, primarily attributed to the presence of SCQDs emitting at higher wavelengths (980-1150 nm).

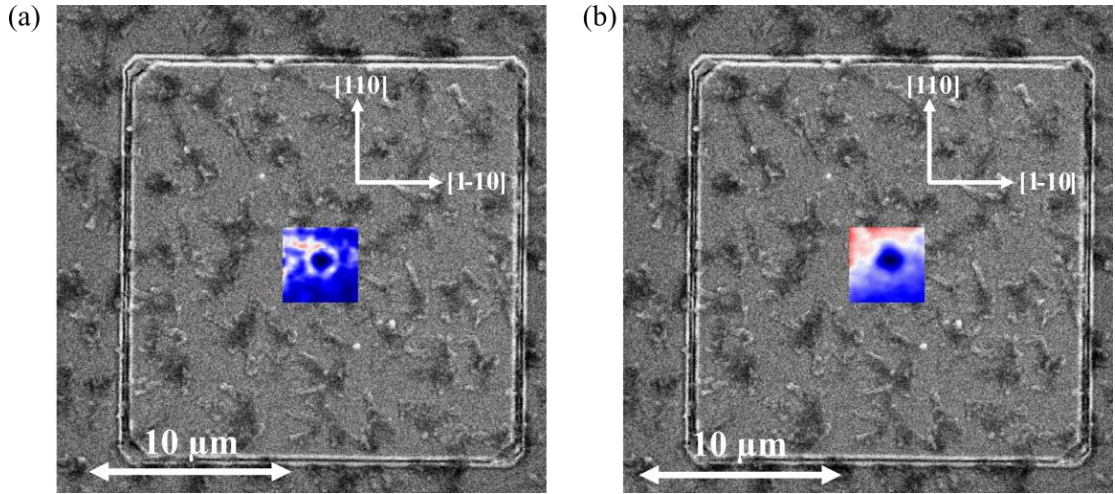

FIG. S2. A CL intensity map of the wetting layer ranging from 885 to 915 nm (a), and the non-positioned QDs between 920 and 980 nm (b), both overlaid with a SEM image of the mesa with the crystallographic orientation of the edges.

The lateral stress arising from the difference in lattice constants between the AlAs and Al$_2$O$_3$ layers propagates to the surface from the mesa, as illustrated in Fig. S3 (a). This leads to tensile strain at the center directly above the aperture, surrounded by a ring of compressive strain, as seen in Fig. S3 (b).

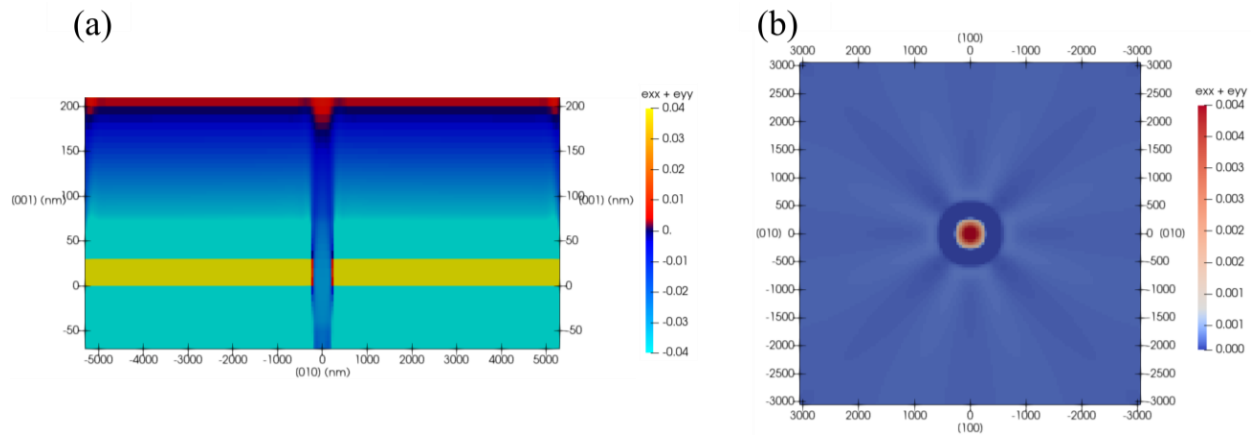

FIG. S3. A side cut through the center of the mesa of the propagation of the strain, $\varepsilon_{xx} + \varepsilon_{yy}$, from the oxidized AlAs layer to the mesa's surface is shown in panel (a) and the surface view of the protruded strain (b). The color scale bar ranges from -4% to 4% in (a) and from 0 to 0.4% in (b). The top surface experiences a tensile biaxial strain induced by the Al$_2$O$_3$ layer, with the largest magnitude directly above the aperture.

Modifying the size of the aperture induces a change in strain strength and profile. During the subsequent growth step, the QD nucleation site, density, and distribution are influenced directly by the applied strain. Fig. S4 (a) displays the calculated surface strain $\varepsilon_{xx} + \varepsilon_{yy}$ and in Fig. S4 (b) we show AFM measurements of four mesas with varying aperture sizes. Note that there is a slight disagreement between the thicknesses used in theory and experiment, which is due to the fact that in theory that is computed as the width of the buried stressor, while in experiment that is deduced from surface AFM scans. Nevertheless, in both theory and experiment. Reducing the AlAs layer width results in smaller nucleation sites with more pronounced site selection, while larger apertures yield a square-shaped distribution of emitters with weaker strain.

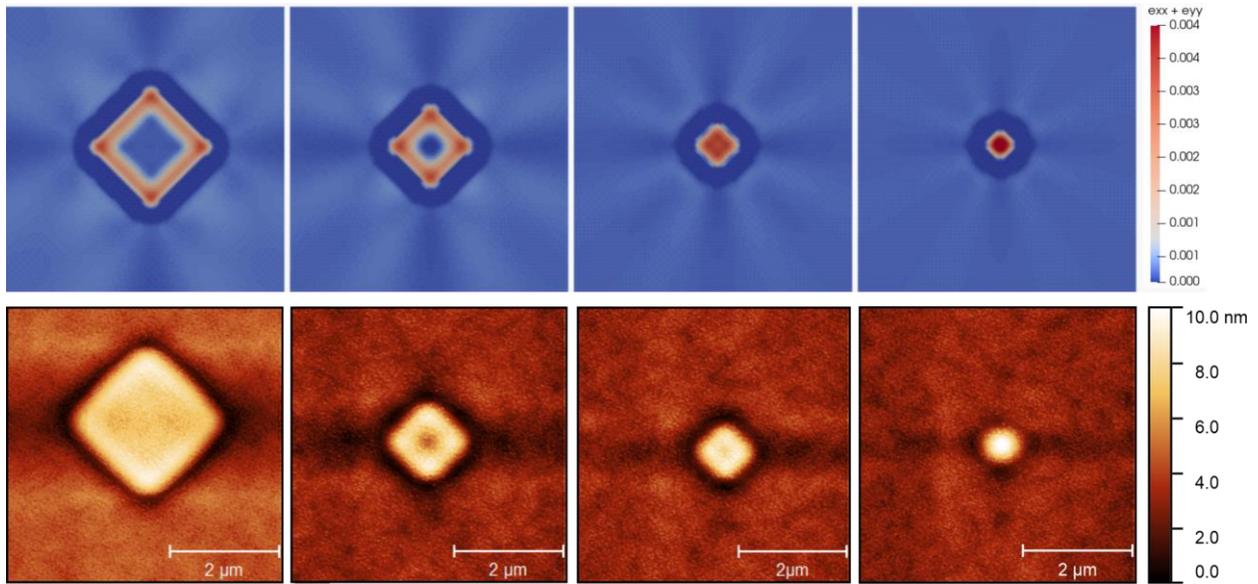

FIG. S4. (a) Computed surface strain $\varepsilon_{xx} + \varepsilon_{yy}$ for aperture widths from left to right: 1500 nm, 1000 nm, 500 nm, 300 nm. (b) AFM images of the overgrown structure at four distinct aperture sizes, spanning from 700 to 3000 nm. Note the slight disagreement in aperture size between theory and experiment. Note that the planar cuts in (a) are rotated by -45 degrees compared to Fig. S3 (b).

The spacing between the aperture and the growth surface of the QDs significantly influences the surface strain, as depicted in Fig. S5 (a). As the aperture-QD spacer thickness increases, the tensile strain at the growth surface also increases, enabling more favorable site-controlled growth of QDs and a more pronounced redshift due to the greater accumulation of indium of the emitter. This effect is particularly pronounced for smaller apertures. Furthermore, as illustrated in Fig. S5 (b), the change in strain intensity between the various layers resulting from the altered distance between the aperture and each respective layer does not impact the density of SCQDs or non-positioned QDs. This allows for a straightforward stacking of multiple layers to enhance optical gain.

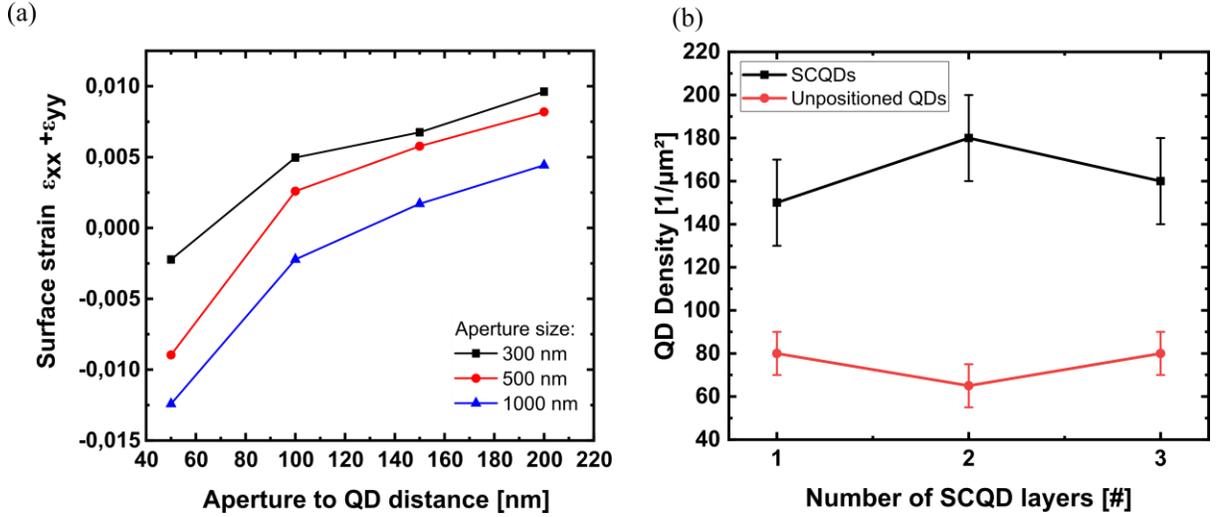

FIG. S5. (a) The surface strain, $\varepsilon_{xx} + \varepsilon_{yy}$, resulting from continuum elasticity theory simulations, is plotted against the separation thickness between the AlAs stressor layer and the QD growth surface. (b) The density of SCQDs and non-positioned QDs versus the number of grown layers.

As a consequence of the strain generated by the buried stressor (illustrated in Fig. S3), QDs located at the growth surface experience tensile strain, leading to a shift in the excitonic emission, as depicted in Fig. S6 (a). The simulated energy shift in QD emission due to the applied surface strain matches the measured emission shift in magnitude. Notably, the extent of applied tensile strain is dependent on the size of the buried stressor, as indicated by the simulation results in Fig. S6 (b). Smaller apertures result in the highest surface tensile strain, with the peak strain occurring precisely at the center of the mesa. Conversely, as the stressor size increases, the strain diminishes sharply, and the strain profile changes as previously mentioned. The strain at the center of the mesa weakens and becomes similar to unstrained GaAs, as the rest of the mesa, while the distribution of strain and QDs adopts a square configuration, as observed in Fig. S4.

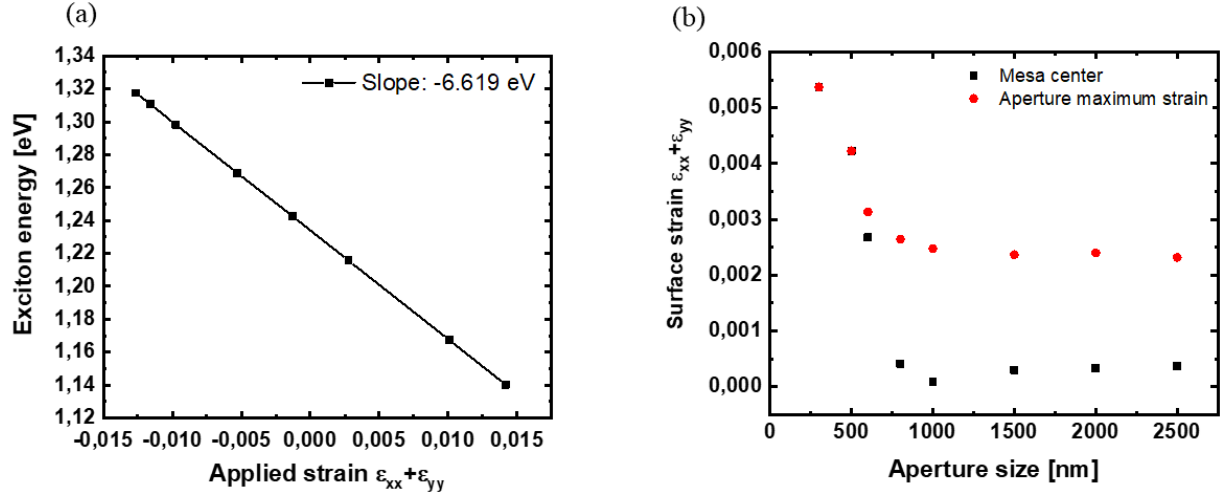

FIG. S6. (a) Excitonic emission of an InGaAs quantum dot (QD) against the applied strain, $\varepsilon_{xx} + \varepsilon_{yy}$, from the aperture, simulated using continuum elasticity theory. (b) The surface strain at the center of the mesa and the highest tensile position on the aperture as a function of the size of the unoxidized AlAs layer.